\documentclass[10pt,journal]{IEEEtran}
\makeatletter
\def\ps@headings{%
\def\@oddhead{\mbox{}\scriptsize\rightmark \hfil \thepage}%
\def\@evenhead{\scriptsize\thepage \hfil \leftmark\mbox{}}%
\def\@oddfoot{}%
\def\@evenfoot{}}
\makeatother
\usepackage{amssymb}
\usepackage{amsfonts,amsmath}
\usepackage[ruled]{algorithm}
\usepackage{algpseudocode}
\usepackage{algorithmicx}
\usepackage{graphicx,epsfig,subfigure}
\usepackage{subfig}
\usepackage{color}
\usepackage[table,xcdraw]{xcolor}
\usepackage{booktabs}

\usepackage{cite}

\newcommand {\mymarginpar}[1]{\marginpar{#1}}
\renewcommand {\marginpar}[1]{}

\def\_{\rule{.3em}{.15ex}}      

\newcommand{\ls}[1]
   {\dimen0=\fontdimen6\the\font
    \lineskip=#1\dimen0
    \advance\lineskip.5\fontdimen5\the\font
    \advance\lineskip-\dimen0
    \lineskiplimit=.9\lineskip
    \baselineskip=\lineskip
    \advance\baselineskip\dimen0
    \normallineskip\lineskip
    \normallineskiplimit\lineskiplimit
    \normalbaselineskip\baselineskip
    \ignorespaces
   }

\newcommand {\bearn}{\begin{eqnarray*}}
\newcommand {\eearn}{\end{eqnarray*}}
\newcommand {\barr}{\begin{array}}
\newcommand {\earr}{\end{array}}

\newcommand {\N}{{\cal N}}

\newtheorem{definition}{Definition}
\newtheorem{property}[definition]{Property}
\newtheorem{proposition}[definition]{Proposition}
\newtheorem{lemma}[definition]{Lemma}
\newtheorem{theorem}[definition]{Theorem}
\newtheorem{corollary}[definition]{Corollary}
\newtheorem{example}[definition]{Example}
\newtheorem{remark}[definition]{Remark}

\newcommand {\benum} {\begin{enumerate}}
\newcommand {\eenum} {\end{enumerate}}

\newcommand {\bdesc} {\begin{description}}
\newcommand {\edesc} {\end{description}}

\newcommand {\bfig}[2] {\begin{figure}
  \centering
  \includegraphics[width=#2]{#1}}
\newcommand {\brotatefig}[2] {\begin{figure}[htbp]
                        \centerline {
                         \epsfig{figure={#1},clip=,angle=-90,width={#2}}}}
\newcommand {\bfigfirst}[2] {\begin{figure}[h]
                        \centerline {
                        \setlength{\epsfxsize}{#2}
                        \epsffile{#1}}}
\newcommand {\efig}[2]{ \caption{#2}
                        \label{fig:#1}
                        \end{figure}
                        \mymarginpar{fig:#1}}
\newcommand {\erotatefig}[2]{ \caption{#2}
                        \label{fig:#1}
                        \end{figure}
                        \mymarginpar{fig:#1}}
\newcommand {\rfig}[1]{Figure \ref{fig:#1}}

\newcommand {\btab}[1]{
                       \begin{table}
                       \centering
                       \begin{tabular}{#1}}
\newcommand {\etab}[3] {
                       \end{tabular}
                       \caption[#3]{#2}
                       \label{tab:#1}
                       \end{table}
                       \mymarginpar{tab:#1}
                       \vspace{.1in}}

\newcommand {\btabular}[1]{\begin{center}
                       \begin{tabular}{#1}}
\newcommand {\etabular}{\end{tabular}
                       \end{center}}

\newcommand {\bdefin}[1]{\begin{definition}
                      \mymarginpar{def:#1}
                      \label{def:#1} }
\newcommand {\edefin}       {\end{definition}}

\newcommand {\bpro}[1]{\begin{property}
                      \mymarginpar{pro:#1}
                      \label{pro:#1} }
\newcommand {\epro}   {\end{property}}

\newcommand {\bprop}[1]{\begin{proposition}
                      \mymarginpar{prop:#1}
                      \label{prop:#1} }
\newcommand {\eprop}       {\end{proposition}}

\newcommand {\blem}[1]{\begin{lemma}
                      \mymarginpar{lem:#1}
                      \label{lem:#1} }
\newcommand {\elem}   {\end{lemma}}
\newcommand {\rlem}[1]{Lemma \ref{lem:#1}}

\newcommand {\bthe}[1]{\begin{theorem}
                      \mymarginpar{the:#1}
                      \label{the:#1} }
\newcommand {\ethe}   {\end{theorem}}
\newcommand {\rthe}[1]{Theorem \ref{the:#1}}

\newcommand {\bproof}{\noindent {\bf Proof.} \ }
\newcommand {\eproof} {\hfill \squares \\ \vspace{.3cm}}
\newcommand {\bcor}[1]{\begin{corollary}
                      \mymarginpar{cor:#1}
                      \label{cor:#1} }
\newcommand {\ecor}   {\end{corollary}}
\newcommand {\rcor}[1]{Corollary \ref{cor:#1}}

\newcommand {\bax}[1]{\begin{axiom}
                      \mymarginpar{ax:#1}
                      \label{ax:#1} }
\newcommand {\eax}       {\vspace{-.1in} \end{axiom}}

\newcommand {\bex}[2]{\vspace{.1in}
                      \begin{example}
                      \mymarginpar{ex:#1}
                       {\bf #2}
                      \label{ex:#1} }
\newcommand {\eex}       {\end{example} \vspace{.3cm} }

\newcommand {\brem}[1]{\begin{remark}
                      \mymarginpar{rem:#1}
                      \label{rem:#1} \em }
\newcommand {\erem}   {\end{remark}}

\newcommand {\beq}[1]{\mymarginpar{eq:#1}
                      \begin{equation}
                      \label{eq:#1} }

\newcommand {\beqno}[1]{\mymarginpar{eq:#1}
                      \begin{eqnarray}
                      \nonumber}

\newcommand {\eeq}       {\end{equation}}
\newcommand {\eeqno}       { && \end{eqnarray}}
\newcommand {\req}[1]{(\ref{eq:#1})}

\newcommand {\bear}[1]{\mymarginpar{eq:#1}
                       \begin{eqnarray}
                       \label{eq:#1} }

\newcommand {\bearno}[1]{\mymarginpar{eq:#1}
                       \begin{eqnarray}
                       \nonumber}

\newcommand {\eear}{\end{eqnarray}}
\newcommand {\eearno}{\end{eqnarray}}
\newcommand {\bsel}{\left \{ \begin{array}{cl}}
\newcommand {\esel}{\end{array} \right.}

\newcommand {\bmat}[1]{\left [ \begin{array}{#1}}
\newcommand {\emat}{\end{array} \right ]}
\newcommand {\bsec}[2]{\mymarginpar{sec:#2}
                       \section{#1}
                       \label{sec:#2} }

\newcommand {\bsubsec}[2]{\mymarginpar{sec:#2}
                       \subsection{#1}
                       \label{sec:#2} }

\def\R{I\kern-0.30em R}
\def\N{I\kern-0.30em N}
\def\P{I\kern-0.30em P}
\newcommand\squares{\vrule height6pt width7pt depth1pt}

\newcommand{\peri}{p}

\begin{document}

\title{PPoL: A Periodic Channel Hopping Sequence with Nearly Full Rendezvous Diversity}

\author{Yi-Jheng Lin and Cheng-Shang~Chang,~\IEEEmembership{Fellow,~IEEE},
\thanks{Y.-J. Lin and C.-S. Chang are with  the Institute of Communications Engineering,
National Tsing Hua University,
Hsinchu 300, Taiwan, R.O.C.
email:   s107064901@m107.nthu.edu.tw, cschang@ee.nthu.edu.tw.}
}
\maketitle

\begin{abstract}
We propose a periodic channel hopping (CH) sequence, called PPoL (Packing the Pencil of Lines in a finite projective plane), for the multichannel rendezvous problem. When $N-1$ is a prime power, its period is $N^2-N+1$, and the number of distinct rendezvous channels of PPoL is at least $N-2$ for any nonzero clock drift. By channel remapping, we construct CH sequences with the maximum time-to-rendezvous (MTTR) bounded by $N^2+3N+3$ if the number of commonly available channels is at least two. This achieves a roughly 50\% reduction of the state-of-the-art MTTR bound in the literature.
\end{abstract}




\bsec{Introduction}{introduction}

The multichannel rendezvous problem that asks two users to meet each other by hopping over their available channels is a fundamental problem in many IoT applications, and it has received a lot of attention lately (see, e.g.,  the excellent book \cite{Book} and references therein).
In the multichannel rendezvous problem, a channel is called a {\em rendezvous} channel of a periodic channel hopping CH sequence  if  the two  users (with any arbitrary clock drift between them) rendezvous on that channel within the period of the sequence. The degree-of-rendezvous (DoR) is the number of {\em distinct} rendezvous channels
within the period of the sequence.
A periodic CH sequence is said to achieve full rendezvous diversity (or maximum rendezvous diversity) for the multichannel rendezvous problem with $N$ channels if its DoR is $N$.


For the multichannel rendezvous problem with $N$ channels, it was shown in Theorem 1 of \cite{Bian2013} that there do not exist deterministic periodic CH sequences that can achieve full rendezvous diversity with periods less than or equal to $N^2$.
The lower bound was further strengthened in Theorem 3 of \cite{DRDS13}:
\begin{eqnarray*}
\peri \ge \left\{\begin{array}{ll}
              N^2+N &\mbox{if $N\le 2$}\\
              N^2+N+1 &\mbox{if $N \ge 3$ and $N$ is a prime power}\\
              N^2+2N & \mbox{otherwise}
              \end{array}
              \right.,
\end{eqnarray*}
where $\peri$ is the  period  of a CH sequence with full rendezvous diversity.

One open conjecture in the multichannel rendezvous problem is whether it is possible to construct a periodic CH sequence that achieves the lower bound.
In the literature, there are various periodic CH sequences that can achieve full rendezvous diversity, see, e.g.,
CRSEQ \cite{CRSEQ}, JS \cite{JS2011}, DRDS \cite{DRDS13},  T-CH \cite{Matrix2015}, DSCR \cite{DSCR2016}, and IDEAL-CH \cite{ideal}.
The asymptotic approximation ratio, defined as the ratio of the period to the lower bound $N^2$ when $N \to \infty$, is
still 2 for IDEAL-CH, 2.5 for T-CH and DSCR, and 3 for CRSEQ and DRDS. Clearly, there is still a significant theoretical gap between the lower bound and the state-of-the-art CH sequences.

In this letter, we tackle such a conjecture from another perspective. Instead of focusing on the constructions of periodic CH sequences with full rendezvous diversity, we consider periodic CH sequences with ``nearly'' full rendezvous diversity.
To formally define the notion of ``nearly'' full rendezvous diversity
for a periodic CH sequence $\{c(t), t=0,1,\ldots, \peri-1\}$ on $N$ channels, we define the quantity $DoR(d)$ as the number of {\em distinct} rendezvous channels between $\{c(t), t=0,1,\ldots, \peri-1\}$ and $\{c(t+d), t=0,1,\ldots, \peri-1\}$. Clearly, a periodic CH sequence achieves full rendezvous diversity if $DoR(d)=N$ for all $d \ne 0$. As such, we call a periodic CH sequence achieves ``nearly'' full rendezvous diversity
if $DoR(d)$ is very close to $N$ for all $d \ne 0$.
For this, we  propose a periodic CH sequence, called {\em PPoL}, from Packing the Pencil of Lines in a finite projective plane. When the number of channels is $N$ with $N-1$ being a prime power, the period $p$ of PPoL is $N^2-N+1$, which is smaller than the lower bound in Theorem 3 of \cite{DRDS13}  for a periodic CH sequence with full rendezvous diversity.
We show that the $DoR(d)$ of PPoL is at least $N-2$ for all $d$.

Another theoretical contribution of the PPoL CH sequence is the reduction of the maximum time-to-rendezvous (MTTR).
In this letter, we consider the symmetric, asynchronous, and heterogeneous setting, where (i) the two users are indistinguishable (and follow the same algorithm to generate their CH sequences), (ii) their clocks may not be synchronized, and (iii) their available channels may be different. However, we assume that the $N$ channels are commonly labeled for these two users.
By channel remapping (for the channels not in the available channel set), we show that the MTTR can be bounded by $N^2+3N+3$ (when $N+1$ is a prime power) if the number of commonly available channels is at least two.
On the other hand, the MTTR bounds of the state-of-the-art algorithms, including ORTHO-CH \cite{ideal},  SRR \cite{Localimprove2018}, and FRCH \cite{ChangGY13},
are at least $2N^2$ when the number of commonly available channels is at least one. This shows that one can achieve a 50\% reduction  of the MTTR bound (for a large  $N$) if we relax the assumption of the number of commonly available channels from one to two.


\bsec{The PPoL CH sequence}{PPoL}


\bsubsec{Difference sets and finite projective planes}{fpp}

To construct CH sequences with full rendezvous diversity, it is known in \cite{Hou2011,DRDS13} that it is equivalent to the problem of packing disjoint difference sets.
Though there are efficient algorithms (see, e.g., \cite{Dsetgen2013,Tan2017}) that can find disjoint perfect difference sets in a periodic sequence, there is no lower bound on
 the number of disjoint perfect difference sets that can be found. In other words, given a specified period $\peri$, in general we do not have a lower bound for DoR.

Instead of packing disjoint difference sets,
the key idea of our construction of the PPoL CH sequence is to pack the pencil of lines in a finite projective plane.
For the letter to be self-contained, we first briefly
review the notions of difference sets and  finite projective planes (even though they have been widely used in the literature, see, e.g., \cite{Hou2011,DRDS13,MOR2014,Grantfree2018,ideal}).

\bdefin{difference}{\bf (Difference sets)}
Let $Z_\peri=\{0,1,\ldots, \peri-1\}$.
 A set $D = \{a_0, a_1, \ldots , a_{k-1}\} \subset Z_\peri$  is called a $(\peri,k,\lambda)$-difference set if for every $(\ell\;\mbox{mod}\;\peri) \ne 0$, there exist
at least $\lambda$ ordered pairs $(a_i, a_j)$ such that $a_i-a_j = (\ell\;{\rm mod}\;\peri)$, where $a_i, a_j\in  D$.
A $(\peri,k,1)$-difference set is said to be {\em perfect} if  there exists
exactly one  ordered pair $(a_i, a_j)$ such that $a_i-a_j = (\ell\;{\rm mod}\;\peri)$ for every $(\ell\;\mbox{mod}\;\peri) \ne 0$.
\edefin

\bdefin{fpp}{\bf (Finite projective planes)}
A finite projective plane of order $m$ is a collection of $m^2 +m+1$ lines and $m^2+m+1$ points such that
\begin{description}
\item[(i)] every line contains $m+1$ points,
\item[(ii)] every point is on $m+1$ lines,
\item[(iii)] any two distinct lines intersect at exactly one point, and
\item[(iv)] any two distinct points lie on exactly one line.
\end{description}
\edefin

In the following, we state some well-known facts for the perfect difference sets and the finite projective planes.
Suppose that $D = \{a_0, a_1, \ldots , a_{m}\}$ is an $(m^2+m+1, m+1,1)$-perfect difference set.
Let $\peri=m^2+m+1$.

\noindent
{\bf (P1):} If for some indices $i,j,k,\ell$,
$$((a_i-a_j)\; {\rm mod}\;\peri)= ((a_k-a_\ell)\;{\rm mod}\;\peri) \ne 0,$$
then $a_i=a_k$ and $a_j =a_\ell$. This is due to the unique difference representation property of a perfect difference set.

\noindent
{\bf (P2):}  A time-shifted version of $D$ is still a perfect difference set. Specifically, let
\beq{timeshift1111}
D_\ell=\{(a_0+\ell)\;\mbox{mod}\;\peri, (a_1+\ell)\;{\rm mod}\;\peri, \ldots , (a_{m}+\ell)\;{\rm mod}\;\peri\},
\eeq
$\ell=0,1,2,\ldots, \peri-1$. Then  $D_\ell$ is an $(m^2+m+1, m+1,1)$-perfect difference set (from the unique difference representation property).

\noindent
{\bf (P3):}
There exists exactly one common element in $D$ and $D_\ell$ for $\ell \ne 0$
such that
$a_i = (a_j+\ell)\;{\rm mod}\;\peri$ (from the unique difference representation property).
Also, as there is exactly  one  ordered pair $(a_i, a_j)$ such that $a_i-a_j = 1$,
 without loss of generality we can assume that $a_0=0$ and $a_1=1$ and order all the elements in $D$ in the increasing order as follows:
\beq{diff1111}
a_0=0 <a_1=1 < a_2 < \ldots, < a_{m} <\peri.
\eeq




\noindent
{\bf (P4):}
Singer \cite{Singer1938} established an important connection between an $(m^2+m+1, m+1,1)$-perfect difference set and a finite projective plane of order $m$.
\begin{description}
\item[(i)] Let
$\{0,1,\ldots, m^2+m\}$ be the $m^2+m+1$ points.
\item[(ii)] Let  $D_\ell$, $\ell=0,1,2,\ldots, \peri-1$
be the $m^2+m+1$ lines.
\end{description}
Then these $m^2+m+1$ points and $m^2+m+1$ lines form a finite projective plane of order $m$.

\noindent
{\bf (P5):}
The $m+1$ lines in the corresponding finite projective plane that contain point $0$ are $D_0, D_{\peri-a_1}, D_{\peri-a_2}, \ldots, D_{\peri-a_m}$.
These $m+1$ lines are called {\em the pencil of lines} \cite{Singer1938} that contain point 0 (as the pencil point). As the only intersection of the $m+1$ lines is point 0,
 these $m+1$ lines, excluding point 0, are  disjoint, and thus
can be packed into $Z_\peri$, i.e.,
$\{D_0, D_{\peri-a_1}^0, \ldots, D_{\peri-a_m}^0\}$ forms a partition of $Z_\peri$,
where
\beq{diff1123}
D_{\peri-a_i}^0 =D_{\peri-a_i}\backslash \{0\}, \quad i=1,2, \ldots, m.
\eeq

\bsubsec{Construction of the CH sequence}{construction}

In this section, we propose the PPoL algorithm for constructing CH sequences with nearly full rendezvous diversity.
For this, one first constructs an $(m^2+m+1, m+1,1)$-perfect difference set, $D = \{a_0, a_1, \ldots , a_{m}\}$ with
  \beq{diff1111c}
a_0=0 <a_1=1 < a_2 < \ldots, < a_{m} < \peri,
\eeq
where $\peri=m^2+m+1$. This is feasible when $m$ is a prime power \cite{Singer1938}.
Then one assigns each channel to a line packed in $Z_\peri$. Specifically,
 the PPoL CH sequence $\{c(t), t=0,1,\ldots, \peri-1\}$ is constrcuted by assigning channel 0 to the time slots in $D_0$ and channel $i$, $i=1,2, \ldots, m$,  to the time slots in $D_{\peri-a_i}^0$, i.e.,
\bear{PFPP1111}
c(t)= \left\{\begin{array}{ll}
              0 &\mbox{if $t\in D_0$}\\
              i &\mbox{if $t \in D_{\peri-a_i}^0$ for some $i \ne 0$}
              \end{array}
              \right..
\eear


\begin{algorithm}\caption{The PPoL CH sequence}\label{alg:PPoL}

\noindent {\bf Input}  A set of $m+1$ channels $\{0,1,2,\ldots, m\}$ with $m$ being a prime power.

\noindent {\bf Output} A CH sequence $\{c(t), t =0,1,\ldots, \peri-1\}$,  where $\peri=m^2+m+1$.

\noindent 1: Let $\peri=m^2+m+1$ and construct a perfect difference set $D = \{a_0, a_1, \ldots , a_{m}\}$ in $Z_\peri$.

\noindent 2: For $\ell=0,1, \ldots ,\peri-1$, let
$$D_\ell=\{(a_0+\ell)\;{\rm mod}\;\peri, (a_1+\ell)\;{\rm mod}\;\peri, \ldots , (a_{m}+\ell)\;{\rm mod}\;\peri\}.$$

\noindent 3: Let $D_{\peri-a_i}^0 =D_{\peri-a_i}\backslash \{0\}$, $i=1,2, \ldots, m$.

\noindent 4: Construct the CH sequence $\{c(t), t =0,1,\ldots, \peri-1\}$
by assigning channel 0 to the time slots in $D_0$ and channel $i$, $i=1,2, \ldots, m$,  to the time slots in $D_{\peri-a_i}^0$

\end{algorithm}

{
	\begin{table*}
		\begin{center}
			\caption{An illustrating example for the PPoL CH sequence with the perfect difference set $D=D_0=\{0,1,4,6\}$ in $Z_{13}$.}
			\begin{tiny}
				\begin{tabular}{|c|c|c|c|c|c|c|c|c|c|c|c|c|c|}
					\hline
					$d=a_k-a_j$           & 0 & 1 & 2 & 3 & 4 & 5 & 6 & 7 & 8 & 9 & 10 & 11 & 12 \\ \hline
					$(k,j)$               &   & (1,0) & (3,2) & (2,1) & (2,0) & (3,1) & (3,0) & (0,3) & (1,3) & (0,2) & (1,2) & (2,3) & (0,1) \\ \hline
					
					$D_0$ shifts $d$         & 0,1,4,6                & {\color{red}1},2,5,7   & 2,3,{\color{red}6},8   & 3,{\color{red}4},7,9    & {\color{red}4},5,8,10   & 5,{\color{red}6},9,11
					& {\color{red}6},7,10,12 & 7,8,11,{\color{red}0}  & 8,9,12,{\color{red}1}  & 9,10,{\color{red}0},2  & 10,11,{\color{red}1},3  & 11,12,2,{\color{red}4}  & 12,{\color{red}0},3,5 \\ \hline
					
					$D_{12}^0$ shifts $d$    & 12,3,5                 & \cellcolor[HTML]{EFEFEF}0,4,6  & 1,{\color{red}5},7   & \cellcolor[HTML]{EFEFEF}2,6,8  & {\color{red}3},7,9  & \cellcolor[HTML]{EFEFEF}4,8,10
					& {\color{red}5},9,11    & 6,10,{\color{red}12}   & \cellcolor[HTML]{EFEFEF}7,11,0 & 8,{\color{red}12},1  & \cellcolor[HTML]{EFEFEF}9,0,2  & 10,1,{\color{red}3} & \cellcolor[HTML]{EFEFEF}11,2,4 \\ \hline
					
					$D_9^0$ shifts $d$    & 9,10,2              & {\color{red}10},11,3  & \cellcolor[HTML]{EFEFEF}11,12,4 & \cellcolor[HTML]{EFEFEF}12,0,5 & \cellcolor[HTML]{EFEFEF}0,1,6 & 1,{\color{red}2},7
					& {\color{red}2},3,8  & 3,4,{\color{red}9}  & 4,5,{\color{red}10}   & \cellcolor[HTML]{EFEFEF}5,6,11  & \cellcolor[HTML]{EFEFEF}6,7,12 & \cellcolor[HTML]{EFEFEF}7,8,0 & 8,{\color{red}9},1 \\ \hline
					
					$D_7^0$ shifts $d$    & 7,8,11   & {\color{red}8},9,12            & \cellcolor[HTML]{EFEFEF}9,10,0  & 10,{\color{red}11},1    & {\color{red}11},12,2    & \cellcolor[HTML]{EFEFEF}12,0,3
					& \cellcolor[HTML]{EFEFEF}0,1,4  & \cellcolor[HTML]{EFEFEF}1,2,5  & \cellcolor[HTML]{EFEFEF}2,3,6   & 3,4,{\color{red}7}      & 4,5,{\color{red}8}      & \cellcolor[HTML]{EFEFEF}5,6,9
					& 6,{\color{red}7},10           \\ \hline
				\end{tabular}
				\label{table:examplePPoL2}
			\end{tiny}
		\end{center}
	\end{table*}
}

In Table \ref{table:examplePPoL2}, we provide an illustrating example for the PPoL CH sequence with the perfect difference set $D=D_0=\{0,1,4,6\}$ in $Z_{13}$. The other three lines that contain point 0 are
$D_{13-1}=D_{12}=\{12,0,3,5\}$, $D_{13-4}=D_9=\{9,10,0,2\}$ and $D_{13-6}=D_7=\{7,8,11,0\}$.
Thus, we have $D_{12}^0=\{12,3,5\}$, $D_9^0=\{9,10,2\}$ and $D_7^0=\{7,8,11\}$. According to Algorithm \ref{alg:PPoL}, we assign channel 0 for $t=0,1,4,6$, channel 1 for $t=3,5,12$,
channel 2 for $t=2,9,10$ and channel 3 for $t=7,8,11$. This leads to the CH sequence
$$\{c(t), t=0,1,\ldots, 12\}=\{0,0,2,1,0,1,0,3,3,2,2,3,1\}.$$
As $D_0$ is a perfect difference set,  channel 0 is a rendezvous channel.
In particular, for $d=1,2, \ldots, 12$, let
$$t_0(d)=1,6,4,4,6,6,0,1,0,1,4,0$$
be the time slots marked in red in the third row of Table \ref{table:examplePPoL2}.
Then
$$c(t_0(d))=c(t_0(d)+d)=0.$$
However, for $d=1,3,5,8,10,12$, channel 1 is not a rendezvous channel. Similarly, for $d=2,3,4,9,10,11$,
channel 2 is not a rendezvous channel. Also, for $d=2,5,6, 7,8,11$, channel 3 is not a rendezvous channel.
As shown in Table \ref{table:examplePPoL2}, for any $d$, there are at most two channels that are not rendezvous channels (see the cells marked in grey).

In the following theorem, we prove the main result of this letter.

\bthe{PPoL} For any $d \ne 0$,
the $DoR(d)$ of the PPoL CH sequence in Algorithm \ref{alg:PPoL} is at least $m-1$ for a system with $m+1$ channels, where $m$ is a prime power.
\ethe

The proof of \rthe{PPoL} requires the following lemma.

\blem{PFPPmain}
If $$d = (a_k-a_j)\;{\rm mod}\;\peri$$ for some $j \ne i$ and $k \ne i$, then there exists $0 \le t \le \peri-1$ such that
\beq{PFPP2222}
c(t)=c(t+d)=i ,
\eeq
i.e., channel $i$ is a rendezvous channel for such a clock drift $d$.
\elem

\bproof
Let $t=(a_j -a_i) \mod \peri$. Since $j \ne i$, $t \in D_{\peri-a_i}^0$. Thus, $c(t)=i$ from \req{PFPP1111}.
On the other hand,
\bearn
&&(t+d) \mod \peri\\
&&=(a_j-a_i +(a_k- a_j)) \mod \peri\\
&&=(a_k -a _i) \mod \peri.
\eearn
Since $k \ne i$, $((t+d) \mod \peri) \in D_{\peri-a_i}^0$. Thus, $c(t+d)=i$ from \req{PFPP1111}.
\eproof

\bproof(\rthe{PPoL})
Since $D$ is a perfect difference set, for any $0< d< \peri$,  there exists a unique ordered pair $(a_j, a_k)$ such that $d= (a_j-a_k)\;{\rm mod}\;\peri$.
It then follows from \rlem{PFPPmain} that except channels $j$ and $k$, every channel is a rendezvous channel when
$d= (a_j-a_k)\;{\rm mod}\;\peri$. Thus, the $DoR(d)$ of the PPoL CH sequence in Algorithm \ref{alg:PPoL} is at least $m-1$ for a system with $m+1$ channels for any clock drift $d$.
\eproof

\bsec{The remapped PPoL CH sequences}{MTTRbound}

In this section, we show how one remaps the PPoL CH sequence for the multichannel rendezvous problem in
the {\em symmetric, asynchronous, and heterogeneous} setting. One popular remapping method is known as {\em random remapping} that randomly re-assigns a channel not in the channel available set to a channel in the channel available set.
As the $DoR(d)$ of the PPoL CH sequence is at least $N-2$ from \rthe{PPoL}, the two users following the PPoL CH sequences with random remapping are guaranteed to rendezvous within $m^2+m+1$ time slots for any prime power $m \ge N-1$ if the number of commonly available channels between these two users is not smaller than three.

Now we show that one can further reduce the requirement for the minimum number of commonly available channels from three to two.
Specifically, for the multiple channel rendezvous problem with $N$ channels, we first construct the
PPoL sequence from an $(m^2+m+1, m+1,1)$-perfect difference set, where $m$ is the smallest prime power not smaller than $N+1$ (instead of $N-1$ in Algorithm \ref{alg:PPoL}). For a user, let $n$ be the number of its available channels, and $n^\prime=m+1-n$ be the number of channels not in its available channels among the $m+1$ channels in Algorithm \ref{alg:PPoL}. Consider the following two cases:

\noindent {\em Case 1.} $n > (N+2)/2$:

In this case, we simply use random remapping.

\noindent {\em Case 2.} $n \le (N+2)/2$:

Since  $m \ge N+1$, we know that
$$n^\prime =m+1-n\ge \frac{N+2}{2} \ge n.$$
Thus, we can remap  the first $n$ channels not in the available channels to the $n$ available channels through a one-to-one deterministic function. For the rest $n^\prime -n$ channels not in the available channels, we
simply use random remapping.

 One key insight of the deterministic remapping in Case 2 is that every available channel of a user is assigned to {\em two lines} of the $m+1$ lines in the PPoL CH sequence.
The detailed steps are shown in Algorithm \ref{alg:PPoLremap}.

\begin{algorithm}\caption{The remapped PPoL CH sequence}\label{alg:PPoLremap}
\noindent {\bf Input}:  A set of available channels ${\bf c}=\{c_0, c_1, \ldots, c_{n-1}\}$ that is a subset of the $N$ channels $\{0,1, \ldots, N-1\}$.

\noindent {\bf Output}: A CH sequence $\{c(t), t=0,1,\ldots \}$ with $c(t) \in {\bf c}$.

\noindent 1: Let $m$ be the smallest prime power such that $m \ge N+1$ and $\peri=m^2+m+1$.

\noindent 2: Use Algorithm \ref{alg:PPoL} to generate the PPoL CH sequence $\{c(t), t =0,1,\ldots, \peri-1\}$.


\noindent 3: Let ${\bf c}^c=Z_{m+1} \backslash {\bf c}= \{c_0^\prime, c_1^\prime , \ldots, c_{m+2-n}^\prime\}$ be the set of channels not in the available channel set.

\noindent 4: {\em Case 1.} $n > (N+2)/2$: Use random remapping, i.e.,
if $c(t)=c_j^\prime$ for some channel $c_j^\prime$ in ${\bf c}^c$, remap $c(t)$ randomly to a channel in ${\bf c}$.

\noindent 5: {\em Case 2.} $n \le (N+2)/2$:
Suppose $c(t)=c_j^\prime$ for some channel $c_j^\prime$ in ${\bf c}^c$. If $j<n$, remap $c(t)=c_j$. Otherwise remap $c(t)$ randomly to a channel in ${\bf c}$.

\end{algorithm}

\bthe{PPoLMTTR}
Consider two users with the available channel sets ${\bf c}_1$ and ${\bf c}_2$.
Suppose that there are at least two commonly available channels between these two users, i.e., $|{\bf c}_1 \cap {\bf c}_2 |
\ge 2$.
 If the two users use the remapped PPoL CH sequences in Algorithm \ref{alg:PPoLremap} to generate their CH sequences, then
these two users are guaranteed to rendezvous within $m^2+m+1$ time slots, where $m$ is the smallest prime power not smaller than $N+1$.
\ethe

As a direct consequence of \rthe{PPoLMTTR}, the MTTR of the remapped PPoL CH sequences in Algorithm \ref{alg:PPoLremap} is $N^2+3N+3$ if $N+1$ is a prime power. Such an MTTR bound is substantially smaller than those from the state-of-the-art algorithms, including $(2N +1)N$ in
ORTHO-CH \cite{ideal}, $(2N +2)N$ in SRR \cite{Localimprove2018}, and $(2N+1)N$ 
in FRCH \cite{ChangGY13}. The reduction of the MTTR bound is due to the assumption that the number of commonly available channels is at least two.

For the proof of \rthe{PPoLMTTR}, we need the following extension of \rlem{PFPPmain}.
The result in \rlem{rendezvouschannel} shows that if a channel of one user appears in the time slots of  two different lines, one from the original assignment and the other from the remapping of a channel not in the available channel set, then that channel is a rendezvous channel
except for one half of the set of clock drifts in \rlem{PFPPmain}.

\blem{rendezvouschannel}
Consider two sequences $\{c_1(t), 0 \le t \le \peri-1\}$ and $\{c_2(t), 0 \le t \le \peri-1\}$.
Suppose that
$c_1(t)=i$ for $t \in D_{\peri-a_{i}}^0$
and that
$c_2(t)=i$ for $t \in D_{\peri-a_{i}}^0 \cup D_{\peri-a_{i_2}}^0$ for some $i_2 \ne i$.
If
\beq{PPoL3300}
d = (a_k-a_j)\;{\rm mod}\;\peri,
\eeq
 for some $k \ne i$,
 there exists $0 \le t \le \peri-1$ such that
\beq{PPoL1111}
c_1(t)=c_2((t+d) \mod \peri)=i .
\eeq
Thus, channel $i$ is a rendezvous channel for such a clock drift $d$.
\elem

\bproof
We consider the following two cases.

Case 1. $k\ne i$ and $j\ne i$:

If $k\ne i$ and $j\ne i$, then by \rlem{PFPPmain}, there exists $0 \le t \le \peri-1$ such that both $t$ and $((t+d) \mod \peri)$ are in $D_{\peri-a_{i}}^0$. Thus $c_1(t)=c_2(t+d)=i$ from \req{PFPP2222}.

Case 2. $k \ne  i$ and $j =i$:

First, we argue that $(2a_i -a_{i_2} -a_k) \mod \peri \ne 0$ when $k \ne i$.
If $(2a_i -a_{i_2} -a_k) \mod \peri =0$, then
\beq{PPoL3311}
(a_i -a_{i_2}) \mod \peri = (a_k -a_{i}) \mod \peri.
\eeq
Since $k \ne i$, $(a_k -a_{i}) \mod \peri \ne 0$.
From the unique difference representation property for  the perfect difference set $D$,
we have from \req{PPoL3311} that $a_i=a_k$ and $a_{i_2}=a_i$. This contracts to the fact that
$k \ne i$.

Since $(2a_i -a_{i_2} -a_k) \mod \peri \ne 0$ for $k \ne i$, there exists a unique pair of $a_{\ell_1}$ and $a_{\ell_2}$ in $D$ such that
\beq{PPoL3322}
(a_{\ell_1}-a_{\ell_2}) \mod \peri =(2a_i -a_{i_2} -a_k) \mod \peri.
\eeq
We argue that $\ell_1 \ne i$ and $\ell_2 \ne i_2$.
If $\ell_1 =i$, then it follows from \req{PPoL3322}
that
\beq{PPoL3355}
(a_{i_2}-a_{\ell_2}) \mod \peri =(a_i -a_{k}) \mod \peri.
\eeq
From the unique difference representation property for  the perfect difference set $D$,
we have from \req{PPoL3355} that $i_2 =i$ and $\ell_2=k$. This contracts to the fact that
$i_2 \ne i$.
On the other hand, if $\ell_2 =i_2$, then it follows from \req{PPoL3322}
that
\beq{PPoL3366}
(a_{\ell_1}-a_{i}) \mod \peri =(a_i -a_{k}) \mod \peri.
\eeq
From the unique difference representation property for  the perfect difference set $D$,
we have from \req{PPoL3366} that $\ell_1=i$ and $i =k$. This contracts to the fact that
$i \ne k$.

Select
\beq{PPoL3344}
t=(a_{\ell_1}-a_i) \mod \peri.
\eeq
Since $\ell_1 \ne i$, we know that $t \in  D_{\peri-a_{i}}^0$ and thus
\beq{PPoL3377}
c_1(t)=i.
\eeq
On the other hand, we have from \req{PPoL3300} and \req{PPoL3322} that
\bear{PPoL3388}
t+d&=&(a_{\ell_1}-a_i)+(a_k -a_i) \mod \peri \nonumber
\\
&=&(a_{\ell_2}-a_{i_2}) \mod \peri .
\eear
Since $\ell_2 \ne i_2$, we know that $t+d \in  D_{\peri-a_{i_2}}^0$ and thus
\beq{PPoL3399}
c_2(t+d)=i.
\eeq


\eproof

\bproof (Proof of \rthe{PPoLMTTR})
Let $n_1=|{\bf c}_1|$ (resp. $n_2=|{\bf c}_2|$) be the number of available channels of user 1 (resp. user 2).
Also, let $n_{1,2}=|{\bf c}_1 \cap {\bf c}_2 |$ be the number of common channels between these two users.
We consider the following two cases:

Case 1. $n_1 + n_2 \ge N+3$:

Since ${\bf c}_1$ and ${\bf c}_2$ are subsets of the $N$ channels,
$$N \ge |{\bf c}_1 \cup {\bf c}_2 |=n_1 +n_2 -n_{1,2}.$$
Since we assume that  $n_1 + n_2 \ge N+3$ in this case, we have $n_{1,2} \ge 3$. As a result of \rthe{PPoL}, these two users are guaranteed to rendezvous within $\peri$ time slots (even without the need of remapping).

Case 2. $n_1+n_2 \le N+2$:

Without loss of generality, we assume that $n_2 \le n_1$. In this case, we have that $n_2 \le (N+2)/2$.
According to Case 2 of Algorithm \ref{alg:PPoLremap},
the first $n_2$ channels  in ${\bf c}^c_2$ are remapped to the $n_2$ channels in ${\bf c}_2$ through a one-to-one deterministic function. In other words, for each available channel $i$ in ${\bf c}^c_2$,
we have $c_2(t)=i$ for $t \in D_{\peri-a_{i}}^0 \cup D_{\peri-a_{i_2}}^0$ for some $i_2 \ne i$.


Since we assume that $|{\bf c}_1 \cap {\bf c}_2| \ge 2$, there exist two {\em distinct} channels
 $\alpha$ and $\beta$  in
 ${\bf c}_1 \cap {\bf c}_2$.
For the CH sequences of these two users $\{c_1(t), 0 \le t \le \peri-1\}$ and $\{c_2(t), 0 \le t \le \peri-1\}$, we know that
$c_1(t)=\alpha$ for $t \in D_{\peri-a_{\alpha}}^0$
and that
$c_2(t)=\alpha$ for $t \in D_{\peri-a_{\alpha}}^0  \cup D_{\peri-a_{\alpha_2}}^0$ for some $\alpha_2 \ne \alpha$.
Similarly,  we also know that
$c_1(t)=\beta$ for $t \in D_{\peri-a_{\beta}}^0$
and that
$c_2(t)=\beta$ for $t \in D_{\peri-a_{\beta}}^0 \cup D_{\peri-a_{\beta_2}}^0$ for some $\beta_2 \ne \beta$.

Now we prove by contradiction that at least one of the two commonly available channels is a rendezvous channel.
Suppose that both channels are not rendezvous channels. Then we have from \rlem{rendezvouschannel} that
the clock drift $d$ must satisfy
$$d=(a_\alpha -a_{j_1}) \mod \peri= (a_\beta -a_{j_2}) \mod \peri$$
for some $j_1$ and $j_2$.
From the unique difference representation property for  the perfect difference set $D$,
we have $\alpha=\beta$ and $j_1 =j_2$. This contradicts to
the assumption that $\alpha\ne \beta$.
\eproof

In \rthe{PPoLMTTR}, we have shown the MTTR bound for the remapped PPoL CH sequence if the number of commonly available channels between the two users is at least two. To further reduce
the number of  commonly available channels to 1, we need to ensure that each available channel of a user is mapped to two lines.
This is stated in the following corollary.

\bcor{PPoLMTTRb}
Consider two users with the available channel sets ${\bf c}_1$ and ${\bf c}_2$.
Let  $m$ be the smallest prime power such that $m \ge N+1$.
 If the two users use the remapped PPoL CH sequence in Algorithm \ref{alg:PPoLremap} to generate their CH sequences and
 \beq{PPoL8888}
\max[n_1, n_2] \le \frac{N+2}{2},
\eeq
then
these two users are guaranteed to rendezvous on every commonly available channel within $m^2+m+1$ time slots.
\ecor

\bproof
According to Case 2 of Algorithm \ref{alg:PPoLremap} and the assumption in \req{PPoL8888},  each available channel of each user is mapped to two lines, i.e., for each available channel $i$ in ${\bf c}_1 \cap {\bf c}_2$,
we have $c_1(t)=i$ for $t \in D_{\peri-a_{i}}^0 \cup D_{\peri-a_{i_1}}^0$ for some $i_1 \ne i$,
and $c_2(t)=i$ for $t \in D_{\peri-a_{i}}^0 \cup D_{\peri-a_{i_2}}^0$ for some $i_2 \ne i$.


Now we argue that each channel $i$ in ${\bf c}_1 \cap {\bf c}_2$ is a rendezvous channel.
This holds trivially for the clock drift $d=0$.
Suppose that channel $i$ in ${\bf c}_1 \cap {\bf c}_2$  is not rendezvous channels for some $d \ne 0$. Then we have from \rlem{rendezvouschannel} that
\beq{PPoL8823}d=(a_i -a_{j_1}) \mod \peri
\eeq
for some $j_1$.
By interchanging $c_1(t)$ and $c_2(t)$ in \rlem{rendezvouschannel}, we also know that
\beq{PPoL8824}d= (a_{j_2}-a_i) \mod \peri
\eeq
for some $j_2$.
From the unique difference representation property for  the perfect difference set $D$,
we must have $i=j_2$ and $j_1 =i$. This contradicts to
the assumption that $d \ne 0$.
\eproof

\bsec{Conclusion}{con}

In this letter, we  proposed in Algorithm \ref{alg:PPoL} the PPoL CH sequence with $DoR(d)\geq N-2$ for the multichannel rendezvous problem with $N$ channels. Such a CH sequence guarantees the rendezvous of two users if
the number of commonly available channels is at least three. The remapped PPoL CH sequence in Algorithm \ref{alg:PPoLremap}  further relaxes the number of commonly available channels from three to two.
These new results have a roughly 50\% reduction of the state-of-the-art MTTR bound in the literature.

Through  extensive simulations, we also observed that the expected time-to-rendezvous (ETTR) of the PPoL CH sequence (with random remapping) is almost the same as those of the simple random algorithm and several existing algorithms, including ORTHO-CH \cite{ideal}, CRSEQ \cite{CRSEQ}, DRDS \cite{DRDS13}, and T-CH \cite{Matrix2015}. Due to space limitations, these numerical results for ETTR's are omitted.



\end{document}